\documentclass[aps,prb,preprint,groupedaddress,floatfix]{revtex4}
\usepackage{graphicx,latexsym}

\begin{document}
\title{Thermal instability of decahedral structures in platinum nanoparticles}
\author{D. Schebarchov}
\affiliation{School of Chemical and Physical Sciences,
Victoria University of Wellington, Wellington 6001, New Zealand}
\author{S. C. Hendy}
\affiliation{School of Chemical and Physical Sciences,
Victoria University of Wellington, Wellington 6001, New Zealand}
\affiliation{MacDiarmid Institute for Advanced Materials
and Nanotechnology, Industrial Research Ltd, Lower Hutt 6009, New Zealand}
\date{\today}
\begin{abstract}
We conduct molecular dynamics simulations of 887 and 1389-atom decahedral platinum nanoparticles using an embedded atom potential. By constructing microcanonical caloric curves, we identify structural transitions from decahedral to fcc in the particles prior to melting. The transitions take place during phase coexistence and appear to occur via melting of the decahedral structure and subsequent recrystallisation into the fcc structure.
\end{abstract}
\maketitle
\section{\label{intro}Introduction}
It has been known for some time that nanoscale particles of fcc metals can adopt non-crystalline structures, such as icosahedra and truncated decahedra, as they trade internal strain for energetically favourable close-packed surface facets \cite{Ino66,Allpress67}. Generally it is found that icosahedra are favoured at small sizes and truncated decahedra at intermediate sizes before the particles eventually adopt their crystalline fcc structures \cite{Marks94}. However, much less is known about the thermal stability of such non-crystalline phases. Molecular dynamics simulations of Lennard-Jones clusters have found that the non-crystalline phases became more stable at temperatures approaching the melting point i.e. the size threshold at which these clusters prefered the non-crystalline phases tended to increase \cite{Doye01}. However recent experiments by Koga et al \cite{Koga04} where free gold clusters were annealed near the melting point found that small particles underwent a transformation from icosahedral to decahedral structures, a finding inconsistent with earlier theoretical considerations.

Other simulations have suggested that solid-liquid phase coexistence may play a role in structural transitions \cite{Hendy05b,Hendy06a}. There are now several instances where structural transitions have been seen in partially melted clusters. In one instance, a transition from an icosahedral to decahedral structure, driven by a preference for the melt to wet the (100)-facets of the truncated decahedron, was observed in a 1415-atom nickel cluster \cite{Hendy05b}. In another instance, a transition from decahedral structures to icosahedral structures was seen in partially melted palladium clusters \cite{Hendy06a}. The underlying cause of this transition was not determined but it was suggested that it may have been due to the preference of recrystallisation kinetics for icosahedral structures after fluctuations in the liquid fraction of the particles. These studies suggest that phase coexistence (and possibly surface melting) may play an important role in structural transitions in nanoparticles. 

HRTEM studies of platinum nanoparticles have found that they showed a preference for fcc structures down to quite small sizes \cite{EPFL1}. Yet molecular dynamics studies using empirical potentials \cite{Baletto02} suggest that non-crystalline structures ought to be stable at low temperatures at least up to sizes of several thousand atoms. This inconsistency between theory and experiment may be due either to the favouring of fcc structures by growth kinetics \cite{Hall97} or to the thermal instability of non-crystalline structures \cite{Baletto05}. Platinum clusters are of particular interest because of their applications to catalysis \cite{Jellinek93,Jellinek95}. Catalytic activity depends critically on the surface structure of the nanoparticle and thus the thermal stability of small fcc and non-crystalline particles is directly relevant to their catalytic efficiency. 

In this paper, we report on molecular dynamics simulations of platinum clusters using an embedded atom method (EAM) potential \cite{Foiles86} where we observe structural transitions from non-crystalline truncated decahedral structures to fcc stuctures prior to melting. As in the case of the transitions seen in Pd clusters \cite{Hendy06a}, the structural stransformations appear to occur via a melting and recrystallisation process although in this case the kinetics are an order of magnitude slower than in Pd.

\section{\label{sec:1}Methodology}   
\begin{figure}
\resizebox{\columnwidth}{!}{\includegraphics{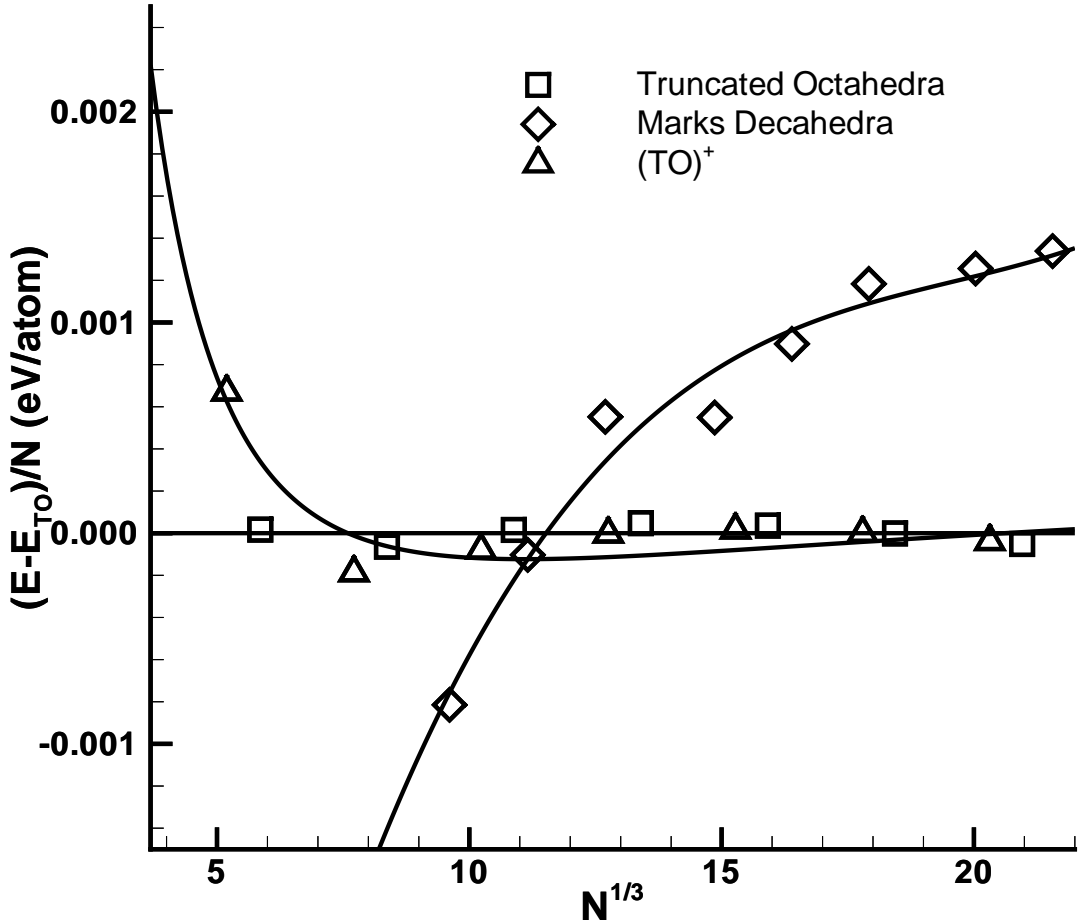}}
%\resizebox{0.55\textwidth}{!}{% Change size with the factor xxx in \resizebox{xxx\te...
%  \includegraphics{test.eps} % replace test.eps with your eps file
%\vspace{5cm}       % Give the correct figure height in cm
\caption{Comparison of the relaxed zero temperature energies per atom of the Marks decahedra, truncated octahedra and the (TO)$^+$ fcc closed-shell sequences for platinum clusters. Energies per atom are given relative to a fit (cubic in $N^{1/3}$) to the energies of the truncated octahedra sequence.}
\label{fig:1}       % Give a unique label
\end{figure}

We use molecular dynamics to simulate the caloric curves of Pt clusters using an EAM potential \cite{Foiles86}. We attempted to identify the stable structures at zero temperature by comparing the potential energies of closed-shell truncated octahedra (including variants such as the (TO)$^-$ and (TO)$^+$ structures \cite{Cleveland99}), cuboctahedra, Mackay icosahedra, and Marks and Ino decahedra \cite{Marks94}. For each structural sequence the potential energy per atom, $E$, was fitted to the expression $E(N) = A + B N^{-1/3} + C N^{-2/3}$. We find that Marks decahedra are favoured at relatively small sizes (887 and 1389-atoms), with fcc structures being favoured at sizes above this. At intermediate sizes the most stable fcc structure is the (TO)$^+$ although at large sizes the truncated octahedra appear to be favoured. In Fig.~\ref{fig:1} we only show the relaxed energies and fits for the the Marks decahedra sequence, the (TO)$^+$ sequence and the truncated octahedra sequence. 

We then construct caloric curves for the lowest energy structure identified for each size range in the constant energy (microcanonical) ensemble using the following procedure: at each fixed total energy the cluster was equilibrated for at least 150000 time steps ($\Delta t = 3.63$  fs) and then the kinetic energy was averaged over a further 150000 to obtain a temperature. An energy increment of 1.0 meV/atom was used to adjust the total energy between each simulation by a uniform scaling of the kinetic energy. To identify and characterise solid-liquid coexistence, we follow Cleveland et al \cite{Cleveland94} in using the bimodality of the distribution of diffusion coefficients to distinguish solid and liquid atoms. This method is discussed in further detail in Ref~\cite{Hendy05c} where it was used to characterise the coexisting solid-liquid states in Ag, Cu and Ni clusters. To characterise solid-solid transitions we have used common-neighbour analysis (CNA) \cite{CNA93} using the classification scheme given in Ref.~\cite{Hendy02}. 
 
\section{\label{sec:2}Results} 

\begin{figure}
\resizebox{\columnwidth}{!}{\includegraphics{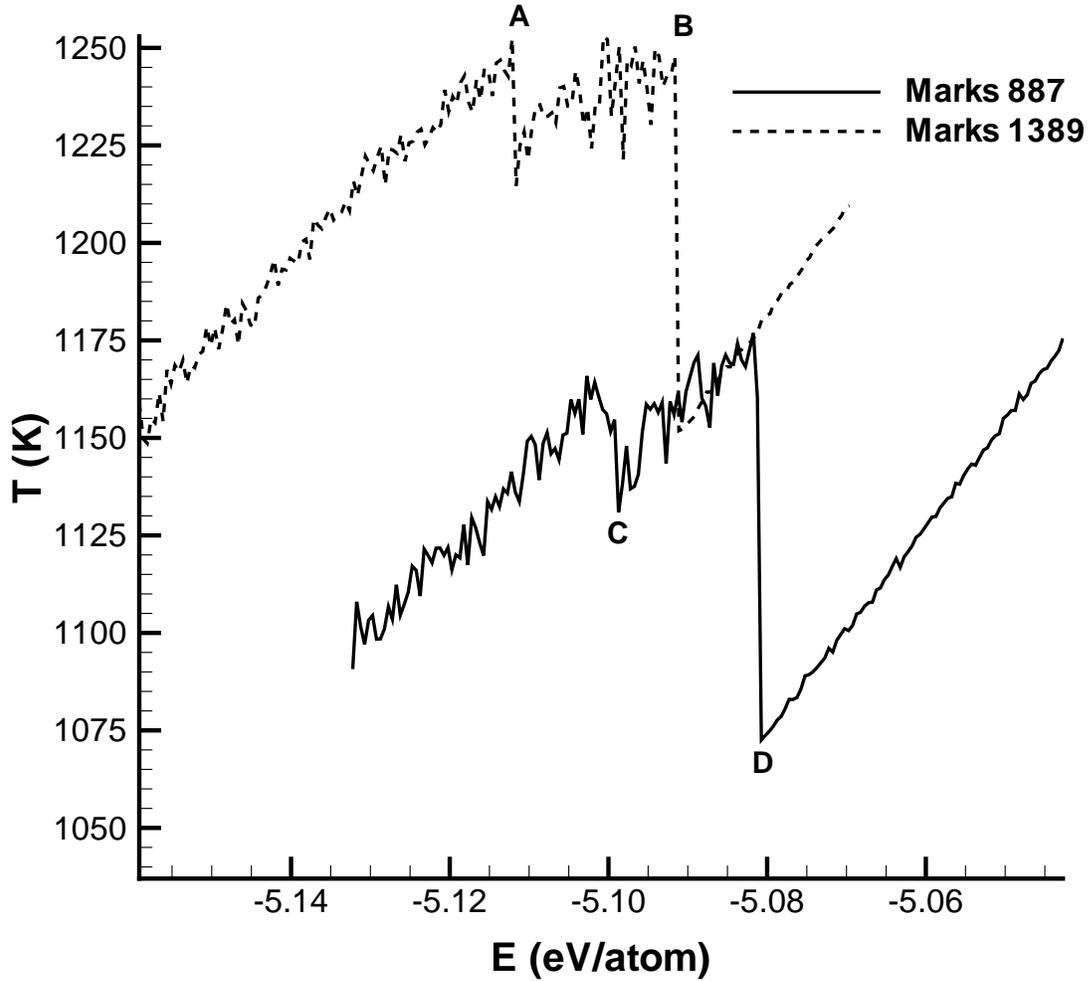}}
\caption{The caloric curves for the 887-atom (solid line) and 1389-atom Marks (dashed line) decahedra. Both curves exhibit a transition during phase coexistence (indicated by the drops in temperature at points A and C respectively) prior to complete melting (which occurs at points B and D respectively.)}
\label{fig:2}       % Give a unique label
\end{figure}

We have examined the thermal stability of the two closed-shell decahedra by constructing microcanonical caloric curves using the procedure described above. Fig.~\ref{fig:2} shows the caloric curves for the 887-atom and 1389-atom Marks decahedra. Note that the curves are qualitatively very similar with two transitions evident in each curve (indicated by the drop in temperature and denoted by A, B, C and D in the figure) over the range of energies considered. The second transition in both cases (at points B and D) is the full melting transition i.e. at energies above the transition point the particles are fully molten. However in both cases the transition prior to melting (A and C respectively) is a structural transition from the decahedral structure to a fcc structure. From the decrease in temperature at the transition (indicating an increase in internal energy) we surmise that the structural transitions are not driven by energetics (i.e. the transitions are not due to a misidentification of the ground state structure, nor the energetic cost of phase coexistence \cite{Hendy05b}).

\begin{figure}
\resizebox{\columnwidth}{!}{\includegraphics{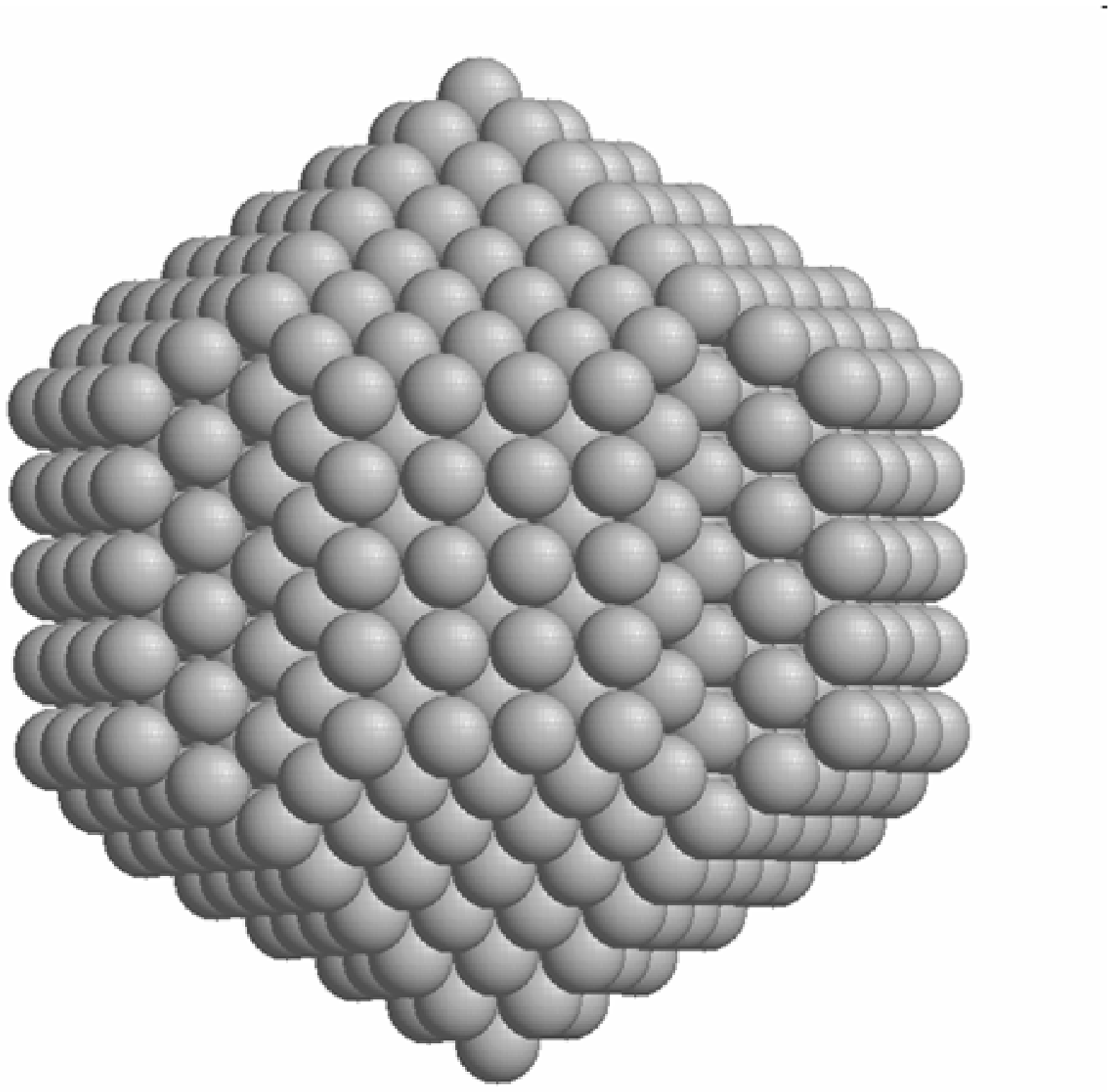} \includegraphics{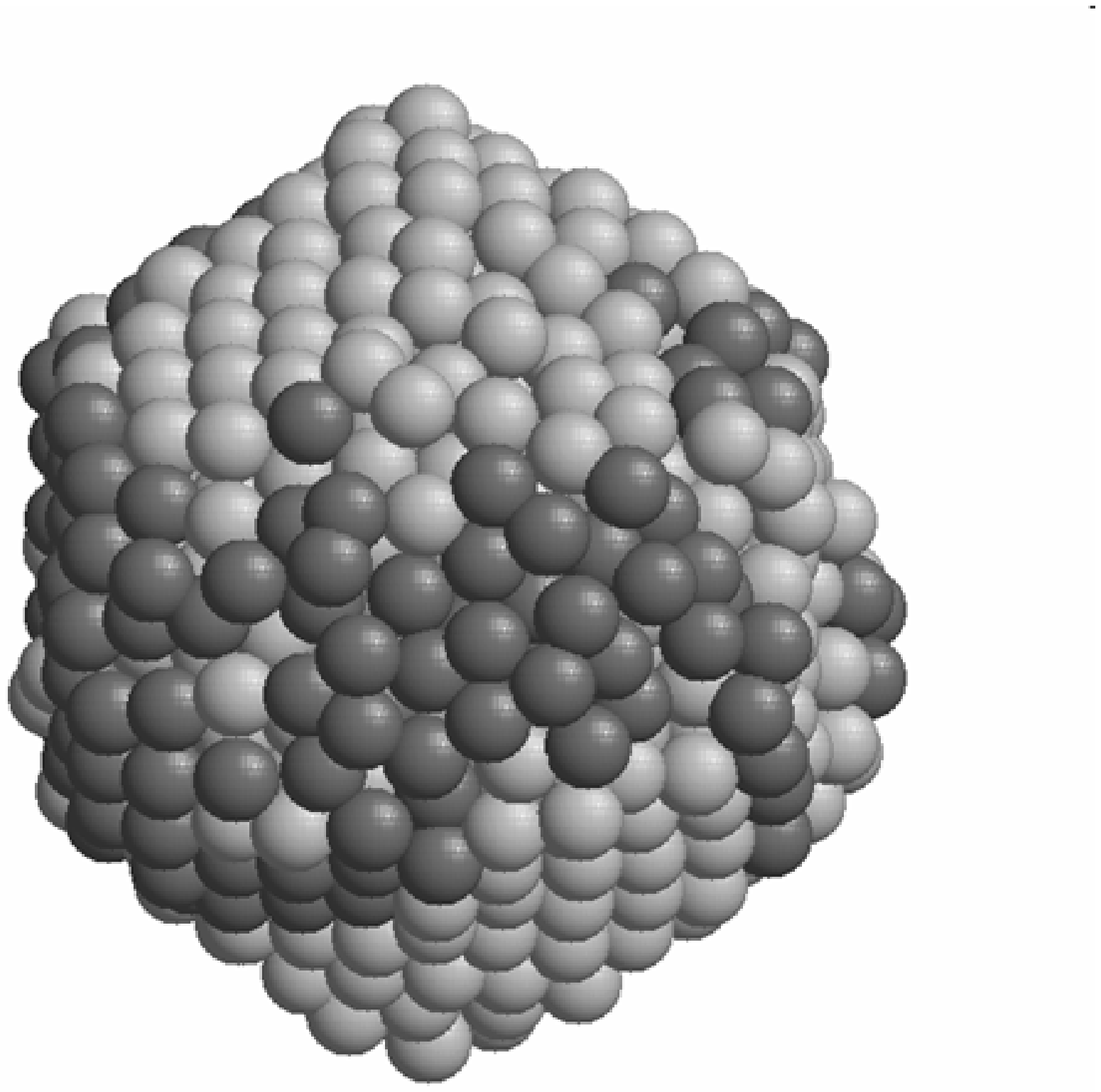} \includegraphics{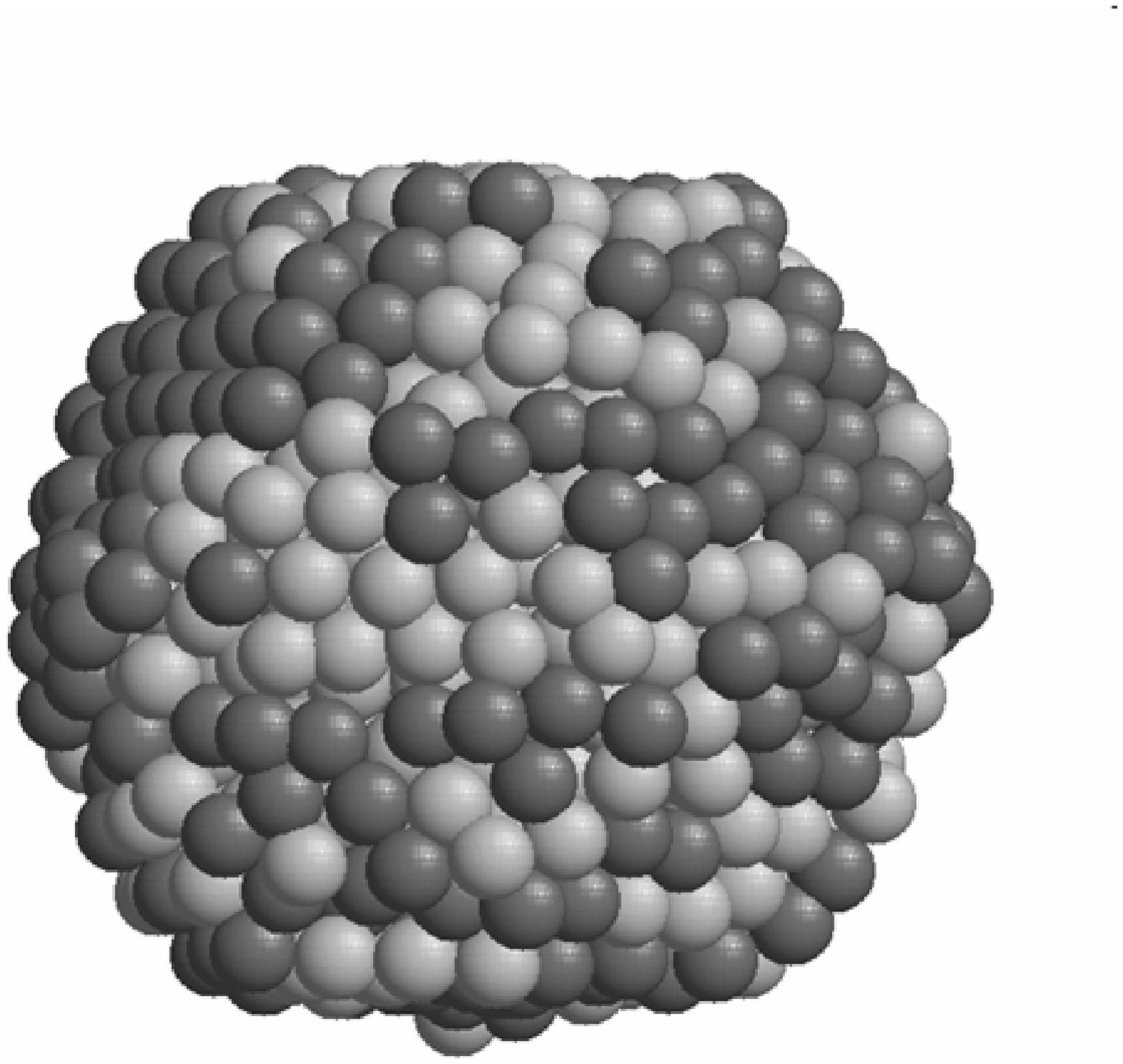}}
\caption{Snap-shots showing the 887-atom particle at 0 K and then immediately prior to and after the transition at C in Fig.~\ref{fig:2}. The solid and liquid atoms, as identified by their mobilities, are shown in light and dark shades respectively.}
\label{fig:3}       % Give a unique label
\end{figure}

We will examine the transition in the 887-atom cluster in more detail (in fact we find similar behaviour in the 1389-atom decahedral cluster). Fig.~\ref{fig:3} shows several snapshots of the 887-atom cluster, firstly at 0 K showing the relaxed Marks decahedral structure, and then immediately prior to and after the transition at C in Fig~\ref{fig:2}. The bimodality of the distribution of mobilities indicates that prior to the transition at C the decahedral structure is in a coexisting solid-liquid state. The location of the liquid atoms in this coexisting decahedral structure (centre frame in Fig.~\ref{fig:3}) suggests that the (100)-facets of the decahedron have melted first. The structure that appears after the transition at C (the right frame in Fig.~\ref{fig:3}) also shows signs of phase coexistence, with liquid atoms covering most of the surface of the cluster. 

\begin{figure}
\resizebox{\columnwidth}{!}{\includegraphics{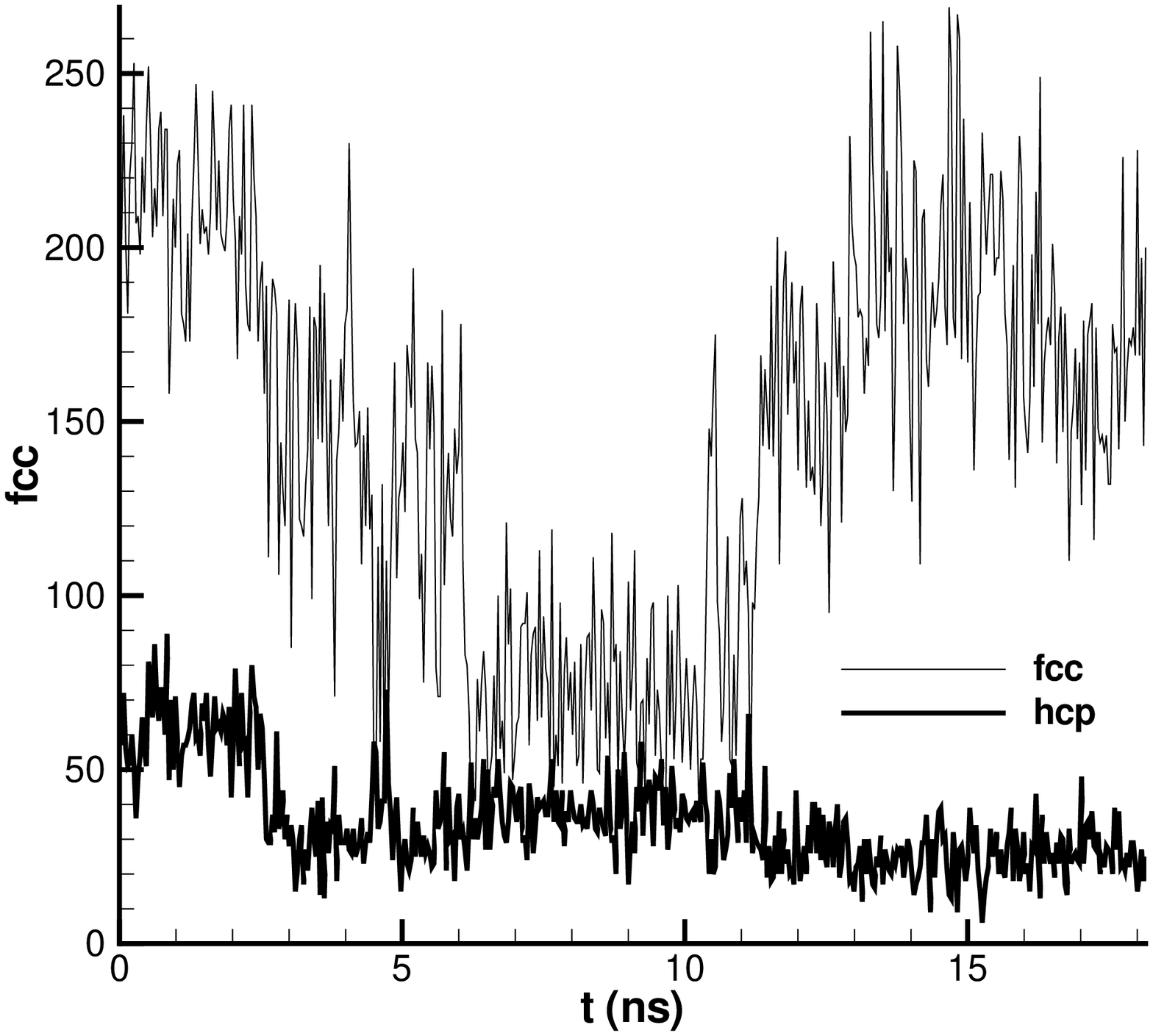}}
\caption{A plot of the number of fcc (light line) and hcp (bold line) atoms of the 887-atom cluster during a long (18 ns) constant energy simulation at $E = -5.108$ eV/atom. The transition from a decahedral structure to an fcc structure appears to occurs via a melting of the decahedra (between $t =$ 2 and 3 ns where the number of hcp atoms drops sharply) followed by a recrystallisation into the fcc structure (between $t =$ 11 and 12 ns).}
\label{fig:4}       % Give a unique label
\end{figure}

\begin{figure}
\resizebox{\columnwidth}{!}{\includegraphics{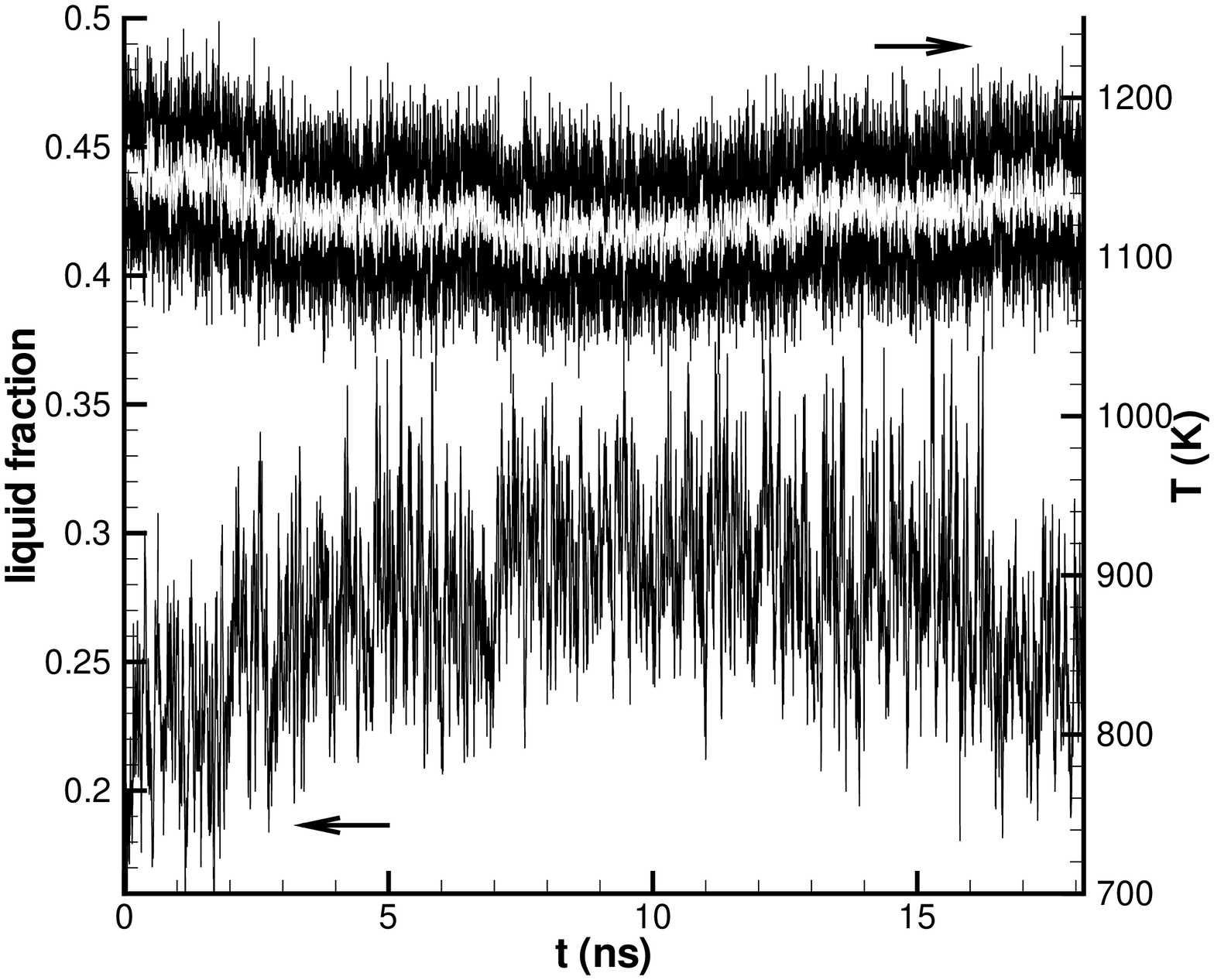}}
\caption{A plot of the instantaneous temperature (plus a time averaged trace of the temperature in white) and liquid fraction of the cluster during the same constant energy simulation at $E = -5.108$ eV/atom as that shown in Fig.~\ref{fig:4}. The liquid fraction peaks at about 0.3 between $t=5-15$ ns before dropping once more as recrystallisation occurs.}
\label{fig:5}       % Give a unique label
\end{figure}

In Fig~\ref{fig:4} we show the results of a CNA structural analysis of a very long duration run constant energy simulation of the 887-atom cluster at $E = -5.108$ eV/atom (an energy just before point C in Fig.~\ref{fig:2}). The cluster is initially in the decahedral structure as can be seen from the high ratio of bulk hcp to fcc atoms identified by the CNA analysis (note that in a relaxed 887-atom Marks decahedron at 0 K our CNA analysis counts 360 bulk fcc atoms and 150 bulk hcp atoms). The transition from this decahedral structure to an fcc structure appears to occurs via a melting of the decahedra between $t =$ 2 and 3 ns where the number of both fcc and hcp atoms drops sharply. Between $t=5$ ns and $t=10$ ns, the number of fcc and hcp atoms remains low indicating considerable disorder. Between $t =$ 11 and 12 ns there is a recrystallisation of the cluster into a fcc structure as seen by the increase in fcc atoms with no corresponding increase in hcp atoms. Fig.~\ref{fig:5} shows the temperature and liquid fraction during the simulation. The liquid fraction peaks at a value of 0.3 at times between the disappearance of the decahedron ($t \sim 3$ ns) and the appearance of the fcc structure ($t \sim 12$ ns). We note also that the changes in temperature correlate well with the changes in liquid fraction, with a minimum in the temperature occuring when the liquid fraction is at its highest.  

\section{\label{sec:3}Discussion}

We have now observed several instances of solid-solid transitions that occur during solid-liquid phase coexistence. In one instance, a 1415-atom Ni icosahedron, a transition was driven by facet-dependent wetting of the melt \cite{Hendy05b}. In a second instance, we observed partial melting of similar sized Pd decahedra followed by recrystallisation into an icosahedral structure \cite{Hendy06a} but this was not driven by the energetics of the of the wetting. The transition reported in this article is most similar to that seen in the Pd decahedra; the decrease in temperature at the transition indicates that it is not energetically driven (unlike the transition in Ni) and it appears to occur as a result of a melting of the decahedral structure followed by recrystallisation into the fcc structure. In Ref.~\cite{Hendy06a} we speculated that the transition was determined by recrystallisation kinetics as icosahedra are often produced by rapid quenches of the melt in small particles. However, the fact that the Pt particles here recrystallise into a fcc structure rather than an icosahedron is not as easily explained by recrystallisation kinetics. It seems likely that this fcc structure is the stable structure at this temperature. We also note that the transition seen in Pd particles occured much more rapidly (recrystallisation occured less than 0.1 ns after the decahedral structure disappeared) than the transition examined here (recrystallisation took 5-10 ns in the 887-atom cluster).

It is not clear what role the phase coexistence plays in the stability of the fcc structure. It may be that the partial melting simply facilitates the kinetics for the solid-solid transition so that it is observable on a timescale accessible by our simulations. On the other hand it may be that the fcc structure is stable only in the presence of the melt. We note that solid-solid structural transitions in coexisting solid-liquid clusters are not ubiquitous in clusters of this size (e.g. see Ref.~\cite{Hendy05c}) but the examples now known from molecular dynamics simulations in Ni, Pd and Pt clusters suggests that such transitions may be reasonably common. While there is no direct experimental evidence for such transitions as yet, it is possible that the icosahedral to decahedral transition in gold clusters observed by Koga et al \cite{Koga04} took place in phase coexisting particles. 

When comparing experimental determinations of nanoparticle structure with corresponding theoretical predictions, one must be aware that particles will frequently be found in kinetically-trapped metastable structures \cite{Hall97}. Growth through atom-by-atom aggregation in an inert-gas aggregation source is known to favour non-crystalline structures in certain circumstances \cite{Baletto05}. Particles may also have undergone melting and freezing during growth, as well as periods of phase coexistence \cite{Kraska04}. In addition if the stable low temperature structure becomes unstable at elevated temperatures it may be difficult to observe because of kinetic trapping. In the case of small Pt nanoparticles, non-crystalline structures have not been observed \cite{EPFL1}. Here we have found a process that favours crystalline structures over non-crystalline structures in Pt particles at elevated temperatures. It is possible that the small fcc particles observed in Ref.~\cite{EPFL1} were kinetically trapped as the result of the high temperature instability of decahedral structures.
 
\section{\label{sec:4}Conclusion}

We have investigated the thermal stability of small decahedral platinum particles. Although these particles appear to be the stable structures at low temperatures, they undergo a transition to an fcc structure at temperatures near the melting point. This instability may be why non-crystalline platinum particles have not been observed experimentally. Previous observations of small platinum particles \cite{EPFL1} may only have identified the metastable kinetically trapped fcc structures.  

\section{\label{sec:5}Acknowledgements}
This work was funded by the Foundation for Research, Science and Technology contract C08X0409 and the MacDiarmid Institute for Advanced Materials and Nanotechnology.
%\begin{figure}
%\resizebox{0.55\textwidth}{!}{% Change size with the factor xxx in \resizebox{xxx\te...
%  \includegraphics{test.eps} % replace test.eps with your eps file
%\vspace{5cm}       % Give the correct figure height in cm
%\end{figure}
%\begin{figure*}
%\vspace*{5cm}       % Give the correct figure height in cm
%\end{figure*}
%\begin{table}
%\begin{tabular}{lll}
%\hline\noalign{\smallskip}
%\end{tabular}
% Or use
%\vspace*{5cm}  % with the correct table height
%\end{table}
%
% BibTeX users please use
% \bibliographystyle{}
% \bibliography{}

\end{document}